Research paper

# Magnetically driven in-plane modulation of the 3D orientation of vertical ferromagnetic flakes.

Hortense Le Ferrand[1]*, Andres F. Arrieta[2]

[1]H. Le Ferrand

School of Mechanical and Aerospace Engineering, 50 Nanyang Avenue, Nanyang Technological University, 639798 Singapore

[2]A. F. Arrieta

School of Mechanical Engineering, Purdue University, 585 Purdue Mall, West Lafayette, IN 47907, United States

**E-mail**: hortense@ntu.edu.sg

**Abstract (200 w)**

External magnetic fields are known to attract and orient magnetically responsive colloidal particles. In the case of 2D microplatelets, rotating magnetic fields are typically used to orient them parallel to each other in a brick-and-mortar fashion. Thanks to this microstructure, the resulting composites achieve enhanced mechanical and functional properties. However, parts with complex geometry require their microstructure to be specifically tuned and controlled locally in 3D. Although the tunability of the microstructure along the vertical direction has already been demonstrated using magnetic orientation combined with sequential or continuous casting, controlling the particle orientation in the horizontal plane in a fast and effective fashion remains challenging. Here, we propose to use rotating magnetic arrays to control the in-plane orientation of ferromagnetic Nickel flakes distributed in uncured polymeric matrices. We experimentally studied the orientation of the flakes in response to magnets rotating at various frequencies and precessing angles. Then, we used COMSOL to model the magnetic field from rotating magnetic arrays and predicted the resulting in-plane orientations. To validate the approach, we created composites with locally oriented flakes. This work could initiate reverse-engineering methods to design the microstructure in composite materials with intricate geometrical shapes for structural or functional applications.

**1| Introduction**

Submitting colloidal particles to external magnetic fields is of interest for studying interparticle interactions,[1] colloidal assembly and patterning,[2,3] as well as for applications such as in health and medicine and robotics.[4–6] More specifically to plate-like particles, also called microplatelets or flakes, controlling their orientation in polymeric matrices is one path to fabricating composites with enhanced mechanical properties and adding functionalities, such as complex shaping and morphing abilities.[7–10] Ferromagnetic particles, such as from iron, cobalt, or nickel, can be oriented and arranged into chains depending on their concentration, the strength of the magnetic field applied, and their dipole-dipole interactions.[11,12] Polymeric composites with anisotropic orientation of ferromagnetic particles were found to exhibit increased mechanical properties, electrical and thermal conductivities along the reinforcement direction, as well as piezoresistivity, electromagnetic shielding, and transparency.[13–16] Although homogeneously orientated composites have their advantage, locally controlling the microstructure over large areas in composites offer substantial potential for engineering hierarchical materials with local properties, for example for touch screens and smart sensors or actuators.[17–19] Also, local microstructure in geometric shapes could help attaining unusual macroscopic properties, such as bioinspired complex crack paths,[20] auxeticity,[21] or shape morphing.[22] However, it remains challenging to purposely control the local orientation of 2D particles in 3D space in composite materials.

Plate-like microparticles are defined in space by two angles $\Phi_p$ and $\theta_p$ (**Figure 1A**). During composite forming, suspended magnetically-responsive microparticles with a sufficient magnetic susceptibility difference with the liquid polymeric matrix orient upon the application of a magnetic field.[23] In the case of ferromagnetic, paramagnetic or diamagnetic flakes, static magnetic fields control the angle $\theta_p$ (**Figure 1B**). To simultaneously set $\Phi_p$ and $\theta_p$, magnetic fields rotating above a critical frequency have been applied on diamagnetic platelets coated with a ferromagnetic material (**Figure 1C**).[24–26] Indeed, magnetically responsive anisotropic particles of microscale dimensions tend to follow the directions of high magnetic fields strength to minimize their magnetic energy. As a result, the particles are found to rotate synchronously with the rotating magnetic field. When the rotation frequency of the magnetic field exceeds a critical value, the motion becomes asynchronous due to the unfavourable damping by the viscous drag. This results in particles aligning biaxially in the plane of rotation of the magnet.[24] The critical frequency $\omega_c$ of the magnetic field rotation depends on the anisotropy in magnetic susceptibility $\Delta\chi = \chi_\parallel - \chi_\perp$ of the particles, with $\chi_{\parallel}$ the susceptibility along the diameter and $\chi_{\perp}$ along the thickness, the viscosity $\eta$ of the surrounding liquid, the magnetic field strength $B_0$, and the Perrin friction factor $\frac{f}{f_0}$:[27]

$$\omega_c = \frac{\Delta\chi \cdot B_0^2}{12 \cdot \eta \cdot \mu_0 \cdot \frac{f}{f_0}}. \qquad \text{(equation 1)}$$

Using rotating permanent magnets, translating solenoids, or solenoids set-ups with on and off periodic control, magnetically responsive flakes have been oriented horizontally and biaxially in so-called brick-and-mortar arrangements ($\Phi_p = cst$ and $\theta_p = 0°$ for all particles) as well as vertically and biaxially ($\Phi_p = cst$ and $\theta_p = 90°$ for all particles) over large areas and in curable polymeric matrices.[28,29] Tuning the orientation of the rotating angle of the magnetic field at defined angles has been used to create similar biaxial orientations at set $\theta_p$ angles and fixed $\Phi_p$ **(Figure 1C)**.[30] Such capability has been utilized to create multilayer composites with set angles $\theta_p$ in each layer.[20,31] These materials are promising for mechanical dampening.[32] However, locally tuning the angles $\Phi_p$ within each layer using magnetic fields is still to be achieved. Realizing it would greatly enhance the geometric design capabilities of reinforced composites.

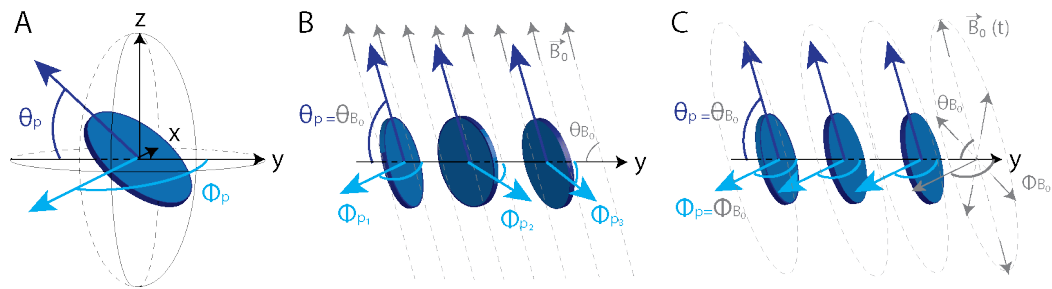

**Figure 1: Orientation of magnetically responsive microparticles under magnetic fields. (A)** Schematics representing a platelet described in the (x-y-z) frame by the two angles $\theta_p$ and $\Phi_p$. **(B)** Schematics representing platelet orientation under static

magnetic fields oriented at $\theta_{B_0}$ fixed. The resulting angles $\Phi_{p_i}$ vary for platelets $i = 1,2, ....$ **(C)** Schematics representing biaxial platelet orientation under magnetic fields rotating in a plane inclined at an angle $\theta_{B_0}$ and at an angle $\Phi_{B_0}$ in the (x-y) plane.

Several methods have been explored to realize in-plane modulation of microplatelet's orientation in liquid matrices, such as acoustophoresis,[33] virtual magnetic molds,[8,34,35] and combinations of 3D printing with magnetic fields.[36,37] Acoustophoresis and virtual magnetic moulds are not practical for the fabrication of composites materials as they require very low viscosity, and 3D printing is still a timely and costly method for large surface areas. One alternative approach potentially reducing processing complexity could be to take advantage of the naturally-occurring modulations of the magnetic field distribution around permanent magnets of diverse shapes or arranged into arrays.[38,39] It has been shown previously that magnetically responsive microplatelets could remain vertically orientated under a rotating magnet, while presenting curved orientations in the horizontal plane in regions located at the edge of the magnet.[40] This feature has been used to tune the curvature of morphing composite bilayers with a leaf-shape geometry.[40,41] Better control of the in-plane orientation of vertical particles (the angle $\Phi_p$) could thus enable more complex composite morphing shapes, or allow geometrical designs with specific, local reinforcement.

In this paper, we explored the orientation control of ferromagnetic nickel flakes experimentally and carry out simple simulations to predict their orientations in selected patterns. The orientation of the Ni flakes was achieved in liquid polymeric matrices which could be turned into a composite after curing. First, we observed the response of the ferromagnetic flakes under vertically rotating magnetic fields. Then, we characterized their response under fields rotating at various precessing angles. Using theoretical and experimental results, we simulated the field lines around the magnets to predict the platelets orientations in the horizontal plane and verified our prediction with magnets of simple geometries. Finally, we built arrays of magnets to achieve unusual radial and circular in-plane orientations of vertical Ni flakes. The results presented show a simple and scalable approach to set the in-plane orientations of vertical ferromagnetic flakes in composites. Such in-plane local variation of the microstructure could enable more geometrical design capabilities. This could be used to create materials adapted to complex external loadings, or with complex deformations or functional response for smart applications.

## 2| Experimental

### 2.1| Materials

Nickel flakes (fine leafing) of diameter 15.8 ± 0.7 μm and thickness 0.58 ± 0.01 μm were obtained from Alfa Aesar, USA. These flakes have non-uniform shapes and an average aspect ratio length over thickness of 27 **(Figure 2)**. The size distributions were obtained from electron micrographs and analysed using Image J and MATLAB. Epoxy was obtained from Hunstmann, Belgium: resin Araldite GY250, hardener Aradur 917 CH and catalyst DY070. Polydimethylsiloxane (PDMS), Sylgard 184 was purchased from Dow Chemicals, USA. Neodymium magnets were used purchased from Supermagnete, Switzerland.

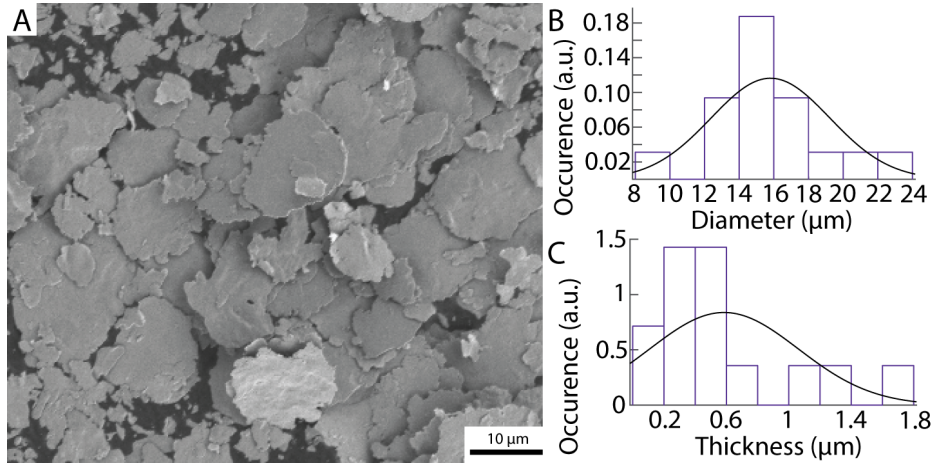

**Figure 2: (A)** Electron micrograph of the Ni flakes. **(B)** and **(C)** are the distributions of flakes diameters and thicknesses, respectively.

## 2.2| Nickel-epoxy composites preparation.

Nickel-epoxy composites were prepared by dispersing 1 vol% of nickel flakes into the epoxy mixture containing 53.5 wt% of the resin, 45.7 wt% of the hardener and 0.8 wt% of the catalyst. The mixture was mixed with an overhead stirrer until homogeneous and degassed under vacuum. The casting was done in a Teflon mould placed over a non-magnetic hot plate set at 180 °C for 4 hours to cure the composite during the magnetic alignment.

## 2.3| Nickel-PDMS composites preparation.

Nickel-PDMS composites were prepared by mixing the PDMS solution at a ratio resin to hardener of 10 to 1 and adding 0.01 vol% of Nickel flakes. A low concentration in Ni powder was used to retain the transparency of the composites after curing. The Ni-PDMS mixture was spread onto an aluminium block and placed under the magnetic setup and above a non-magnetic hot plate set at 120°C for 30 minutes to cure the matrix. For the Ni-PDMS composites used for the determination of the critical frequency, no temperature was applied to avoid curing and to maintain the viscosity constant.

**2.4| Magnetic setups.**

For the magnetically oriented epoxy specimens, the magnet used had dimensions of 5*5*1.5 cm$^3$ and was placed at around 10 cm from the top of the sample. This led to a magnetic field strength at the position of the sample of around 3 mT. A rotating magnetic field was achieved by attaching the magnet to a stirrer rotating at a frequency above 1 Hz. The magnetic field was maintained during the entire curing time of the samples. For the magnetically oriented PDMS specimens, the utilized magnets had dimensions of 0.5*0.5*0.5 cm$^3$ and were mounted onto other stirrers in similar configurations during the entire curing of the composite, at 1.5 cm to the surface of the sample. The magnetic field strength at the position of the sample was also of around 3 mT.

**2.5| Characterization methods.**

Electron micrographs of the raw powder and of the cross-sections of epoxy composites were taken using a scanning electron microscope (Leo 1530, Zeiss, Germany), after sputtering with Pt. The epoxy composites were brittle and could be fractured for the direct observation of the cross-section. For observations in the plane, the samples were polished using sandpaper. The orientation angles of the flakes were measured using the Image J software (NIH, USA). The PDMS composites were observed by optical methods thanks to their transparency using a common digital camera. The critical alignment frequencies under magnets rotating with precessing angles were assessed by recording videos of the mixture under the alignment setup. When the flakes are rotating, a colour change could be seen at the surface of the sample, accompanied by waves (see **SI movie S1**). Flakes rotating asynchronously with the rotating magnetic fields appear to be still in the sample's surface (see **SI movie S2**). The determination of the critical frequencies was done by increasing the frequency every 5 minutes until no more surface change was visible. Indeed, the expected time to have alignment of the nickel flakes under 3 mT was calculated to be of 3.3 s for the epoxy matrix and of 0.9 s for the PDMS using the following equation and the values in **Table 1**:

$$t = \frac{9\pi^2 \cdot \frac{f}{f_0} \eta (\chi_{ps}+1)}{\mu_0 \chi_{ps}^2 B_0^2},$$ (equation 2)

with $\mu_0$ the magnetic susceptibility of vacuum, $\frac{f}{f_0}$ the Perrin friction factor, $\chi_{ps}$ the magnetic susceptibility of the particles.[36]

**Table 1:** Values of experimental parameters.

| Parameter | Symbol | Value | Unit |
| --- | --- | --- | --- |
| Flake diameter | $2 \cdot b$ | 15.8 | µm |
| Flake thickness | $2 \cdot a$ | 0.58 | µm |
| Aspect ratio | $p$ | 27 | - |
| Perrin friction factor | $\frac{f}{f_0}$ | 12.1 | - |
| Magnetic field strength | $B_0$ | 2388 | A/m |
| Magnetic susceptibility | $\chi_{ps}$ | 600 | - |
| Viscosity (epoxy) | $\eta$ | 13.3 | kg/m.s |
| Viscosity (PDMS) | $\eta$ | 3.5 | kg/m.s |

**2.6| Simulations.**

Magnetic fields around permanent magnets were simulated using COMSOL Multiphysics, Version 4.4. The geometries of the magnets were drawn, and the neodymium material was added to the materials library using a relative permeability of

1.05. The material around the magnet was set as air (relative permeability of 1). The magnetic fields were computed using the AC/DC module with magnetic fields, no currents interface. The boundary conditions were set such as:

- The magnetic field was tangential to the boundary on planes of symmetry.
- The magnetic field was perpendicular to the boundary on planes of antisymmetry.

The magnetic field at the surface of the magnet was set at a value of 199'000 kA/m = 2500 Oe, corresponding to the 250 mT measured at the surface of the magnets. The simulation yields the magnetic flux density vectors and streamlines as output, which were plotted in 3D around static permanent magnets. To obtain the projection of the field lines in sample located in a plane at a distance $d$ to the magnet, a time-space transformation was applied (**Figure 3**). This transformation consisted in summing the vectors at each point of space positioned on a circle around the magnet: instead of fixing the point of space and rotating the magnet, we fixed the magnet and rotated the point of interest.

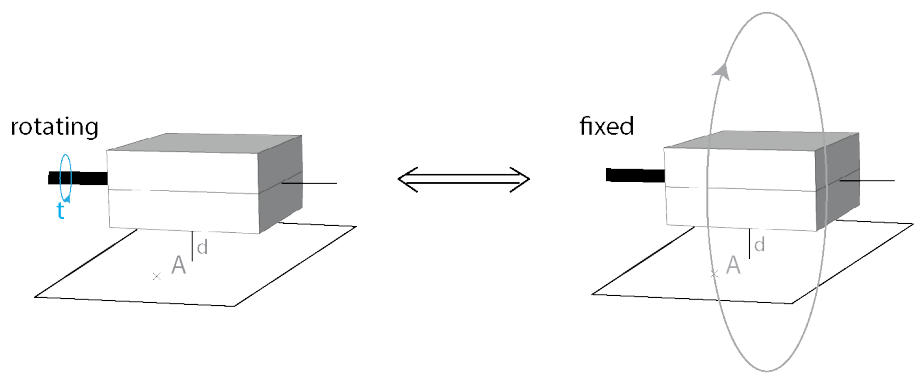

**Figure 3**: Schematics representing the time-space transformation applied.

**3| Results and discussion**

## 3.1| Response of Ni flakes under vertically rotating magnetic fields ($\theta_{B_0} = 90°, \Phi_{B_0}$ variable) and ($\theta_{B_0} = 90°, \Phi_{B_0} = cst$).

Although rotating magnetic fields have been largely applied to orient magnetically responsive para or diamagnetic anisotropic platelets,[24–26,28,42] the response of ferromagnetic flakes to dynamic fields has been overlooked. Therefore, the response of Ni flakes to vertical static and rotating magnetic fields was first observed in epoxy composites containing ca. 1 vol% of flakes (**Figure 4**).

The magnetic fields strength applied were low, below 3 mT, to avoid the formation of large magnetic gradients and the attraction of the Ni flakes to the magnet. Under static vertical magnetic fields ($\theta_{B_0} = 90°$), the Ni flakes align uniaxially ($\theta_p = 90°$, variable $\Phi_p$) (**Figure 4A**). To determine the critical frequency $\omega_c$ of the rotating field to obtain vertical biaxial alignment of the flakes ($\theta_p = 90°$, $\Phi_p$= cst for all particles), the theory for biaxial alignment of superparamagnetic platelets under dynamic rotation fields was applied to ferromagnetic Ni flakes.[24] Since the magnetic susceptibilities parallel and perpendicularly to the long axis $\chi_\parallel$ and $\chi_\perp$ of the flakes are not known, the flakes were assimilated to diamagnetic particles with a thick paramagnetic shell. The critical frequency $\omega_c$ was then estimated using the values presented in **Table 1** and the following formula:

$$\omega_c = \frac{\mu_0 \chi_{ps}^2 B_0^2}{18 \frac{f}{f_0} \eta (\chi_{ps}+1)}, \qquad \text{(equation 3)}$$

with $\mu_0$ the magnetic susceptibility of vacuum, $\frac{f}{f_0}$ the Perrin friction factor, $\chi_{ps}$ the magnetic susceptibility of the particles, and $B_0$ the magnetic field strength. With this approximation, a critical frequency of 0.23 Hz was determined for our epoxy matrix, and of 0.9 for PDMS. Applying a vertical magnetic field ($\theta_{B_0} = 90°, \Phi_{B_0} = 90°$) rotating above 1 Hz, vertical biaxial orientation of the flakes was also observed ($\theta_p =$

90°, $\Phi_p = 90°$) (**Figure 4B**). However, in contrast to superparamagnetic platelets, the ferromagnetic flakes also formed long chains that oriented along the field lines ($\Phi_p = 90°$) due to their dipole-dipole interactions.[12,43] We could observe this chain formation optically in dilute Ni-PDMS transparent films. Chain formation is ubiquitous for ferromagnetic powders exposed to magnetic fields and depends upon filler concentration and magnetic field.[11] The absence of clear chain formation under static magnetic fields might be related to the high viscosity of the matrix and its curing, which did not allow sufficient time for the flakes to translate and self-assemble. However, under rotating fields the flakes rotate and can act as local micromixers, reducing the local viscosity of the matrix.[44] Also, the rotation of the flakes in the (x-y) plane (angle $\Phi_p$) made the observation more difficult since the chains would have formed along the vertical direction.

In addition to the orientation of the flakes under the vertical magnets, we inspected the flake orientations variation along the axis $y$, where $y = 0$ corresponds to the center of the rotating magnet (**Figure 4C**). The angles $\theta_p$ of the flakes decreased as $y$ increased until the edge of the magnet at $y = 2.5$ cm. The flakes are thus sensitive to the field streamlines orientations.

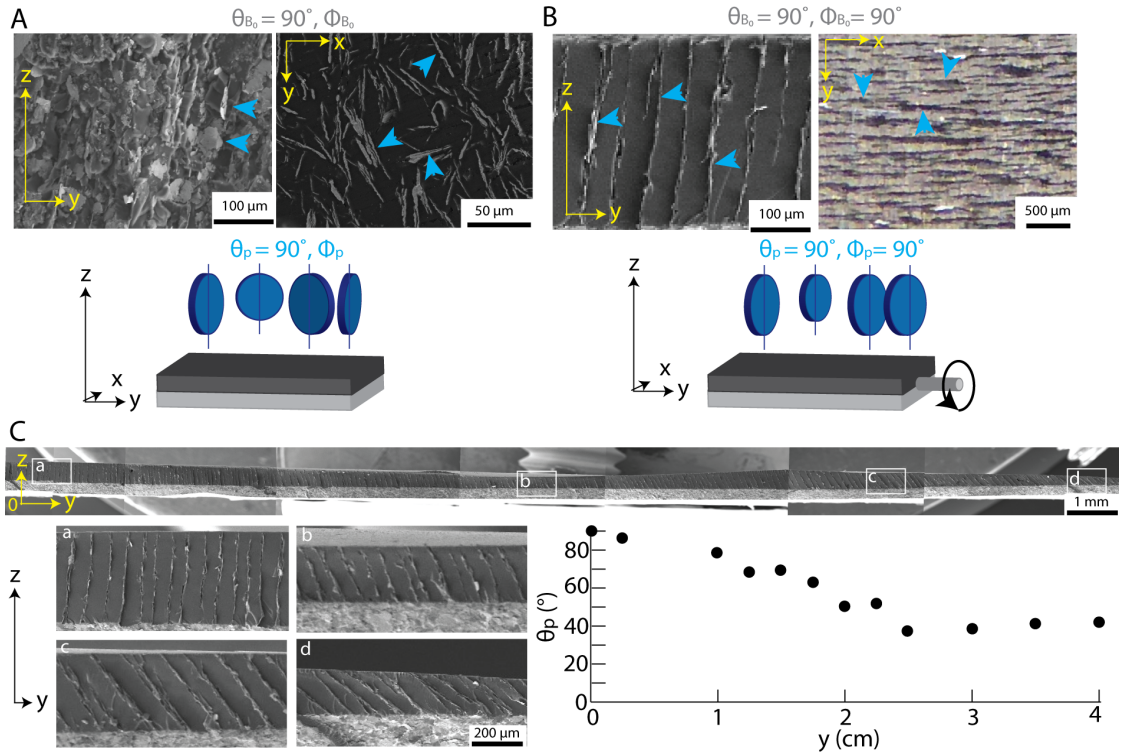

**Figure 4: Response of Ni flakes to vertical magnetic fields. (A)** Electron micrographs of Ni-epoxy composites showing uniaxial orientation under static vertical magnetic field. The blue arrows indicate some Ni flakes. **(B)** Micrograph (left) and optical image (right) of epoxy and PDMS composites, respectively, showing biaxial orientation and chain formation of the Ni flakes under rotating vertical magnetic field. **(C)** Juxtaposed micrographs of the cross section of a large epoxy composite prepared under rotating vertical magnetic field ($\theta_{B_0} = 90°, \Phi_{B_0} = 90°$) centered at $x = 0$ and $y = 0$. Inserts are close-up view. Graph of the particle angle $\theta_p$ as a function of their position along $y$ for $x = 0$.

### 3.2| Response of Ni flakes under magnetic fields rotating with a precessing angle ($\theta_{B_0} = 90° - \alpha, \Phi_{B_0} = cst$).

To further study the response of the Ni flakes to rotating magnetic fields, a precessing angle $\alpha$ was added to the rotating magnet. The in-plane orientations of the flakes $\Phi_p$ in transparent PDMS composites were characterized optically (**Figure 5**). In this configuration, the magnetic field rotates at angles $\theta_{B_0} = 90° - \alpha$ (**Figure 5A**). Previous works on ellipsoid magnetic particles under precessing magnetic fields have reported that biaxial alignment could also be achieved, at a critical frequency that decreased with the precessing angle.[27] Recording videos of the liquid PDMS-Ni mixtures under the precessing magnetic fields, the critical frequency for the alignment could be determined by optical visualization. The experimental results show good agreement with the expected prediction from ref[27], with the observation of an asynchronous regime where the particles align biaxially above a set frequency that decreases with the precessing angle (**Figure 5B,C**). The critical frequency was almost multiplied by 3 at low angles. This is higher than what has been reported elsewhere, where the frequency doubled only.[27] One reason could be that by tilting the magnet without moving its rotation axis or the sample, the magnetic field at the place of observation also increases. Indeed, the magnetic field strength is usually higher at the edge of the magnets. A high magnetic field strength leads to a higher magnetic response of the particles.

The response of Ni flakes to vertical magnetic fields rotating with a precessing angle was thus vertical orientation and biaxial alignment ($\theta_p = 90°, \Phi_p = 90°$) for all precessing angles except 90° (**Figure 5D**). The flakes aligned in the plane along the field lines of the magnet. At $\alpha = 90°$, the magnetic field is horizontal at the position of the sample, below the magnet. This resulted in an increase in transparency of the composite sample and the absence of visible chains (**Figure 5E**). This visual appearance was attributed to the horizontal orientation of the flakes in the plane. Indeed, the field lines are rotating in-plane.

Under rotating vertical magnetic fields, the ferromagnetic flakes align parallel to the plane of rotation described by the magnetic field. To obtain various orientations $\Phi_{pi}$, with i each individual flake, within the (x-y) plane of the sample, it is therefore necessary to have rotating magnetic fields with locally oriented field lines.

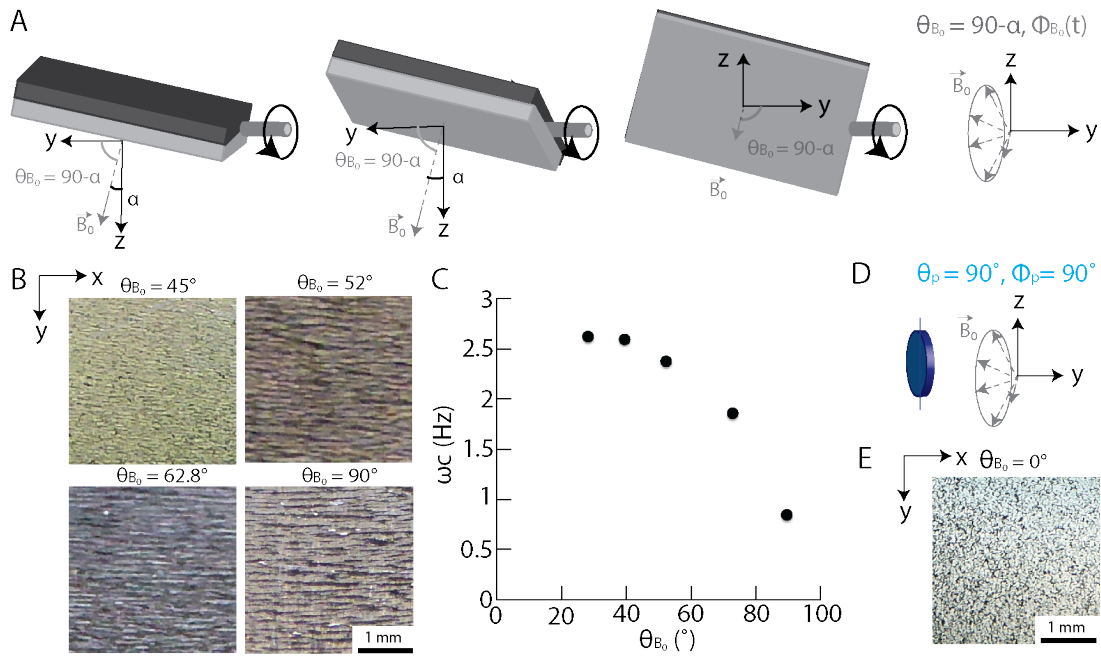

**Figure 5: Nickel flake response to precessing vertical magnetic fields. (A)** Schematics for a magnet rotating with a precessing angle α. **(B)** Optical images in the (x-y) plane of the sample centred below the magnet showing the line formation and alignment of the Ni flakes within the PDMS matrix. **(C)** Critical frequency as a function of the magnetic field angle $\theta_{B_0}$. **(D)** Schematics of the flake orientation with the rotating magnetic field. **(E)** Optical image of the Ni-PDMS composite sample under a precessing angle $\alpha = 90°$ ($\theta_{B_0} = 0$).

**3.3| In-plane modulation of Ni flakes orientation under rotating magnets ($\theta_{B_0} = 90°, \Phi_{B_0}$ = locally varying).**

To modulate the in-plane angle of the flakes while remaining vertical, we looked at their orientation away from the centre of the magnet, where the field lines are oriented along various directions. First, we selected a square and a spherical magnet and used COMSOL to obtain the magnetic streamlines around the magnets. Then, we applied a time-space transformation to mimic the rotating field and obtain the map of $\Phi_{B_0}$ in the (x-y) plane of the sample (**Figure 6**). The vertical orientation of the flakes is assured by having the magnet positioned and rotating above the samples. However, tilting of the flakes could happen at the edge of the magnet due to decreased magnetic force. Ni-PDMS composites were then prepared under the same magnetic setups rotating at a frequency higher than 3 Hz to ensure the asynchronous biaxial alignment of the Ni flakes (**Figure 6A,B**). The experimental specimen exhibited similar flake orientations as predicted by the model. Magnetic field gradients in the sample could give rise to variations in flakes concentration. Although the model is simplistic and its resolution is not yet high enough to match the exact orientation at the microscale, it gives a qualitative view of the expected orientations of the flakes in-plane. We used similar approach to generate other types of patterns using more complex magnetic arrays (**Figure 7**). Small square permanent magnets were assembled to generate more complex magnetic streamlines. The assembly of the magnets was inspired from that of Halbach arrays and others. The intention was to obtain unusual flake orientation in-plane, such as radial and circular (**Figure 7A,B**), extended to large areas (**Figure 7C**), and with local patterns (**Figure 7D**). Although there are variations in platelets concentrations, the composite films with various in-plane flake orientations were obtained.

A more refined model of the magnetic fields and of the flake response to it would push the approach proposed here to a next level. For this, modelling based on the constitutive equations of magnetic fields and with finite element modelling methods could be envisaged.[45,46] With such accurate modelling schemes, reverse engineering of flake orientations in 3D might be realized. The modelling could also include concentration gradients. Reverse-engineering could work in the following way: first, for a given in-plane desired alignment pattern, a magnetic field will be derived. The magnetic field will then be approximated by combinations of magnets or magnets with complex shapes. Then, the time-space transformation will be applied to visualize the in-plane orientation map. Finally, the prediction will be applied in experiments using the set magnetic set-up.

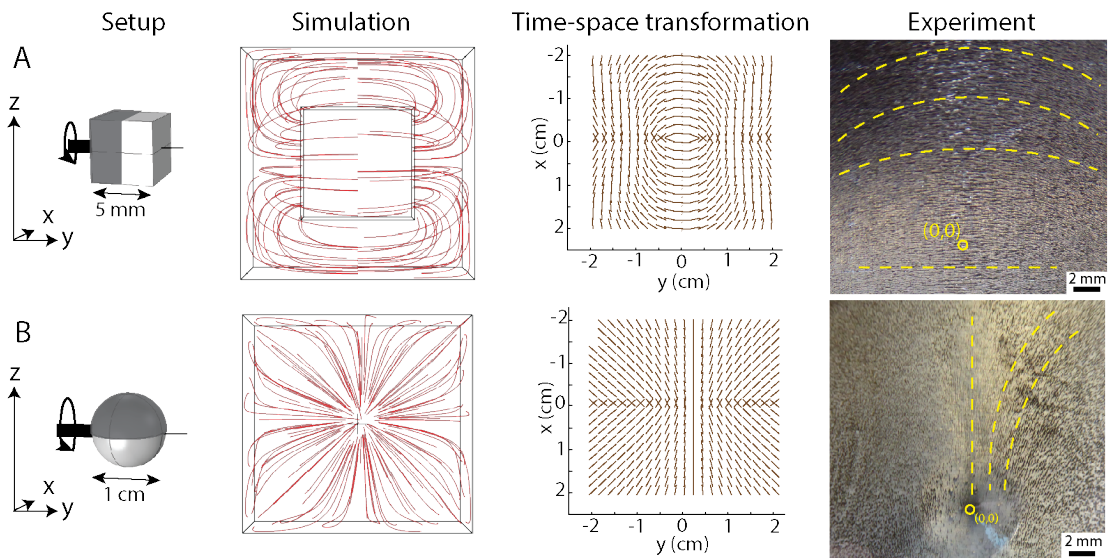

**Figure 6:** Predicting the in-plane modulation of Ni flakes orientation under vertical rotating square **(A)** and spherical **(B)** magnets. Setup, simulation, time-space transformation, and experiment flow to predict and realize in-plane modulation of the $\Phi_p$. The axis of rotation of the magnets is highlighted in bold and the north and south pole in dark and light grey colour, respectively. The red lines around the magnets are the simulated magnetic streamlines, top view in (x-y) plane. The lines in the x-y plot are tilted at the local angles $\Phi_p$ of the flakes. The experimental images are pictures of Ni-PDMS transparent samples after curing. The yellow lines are guides to the eye.

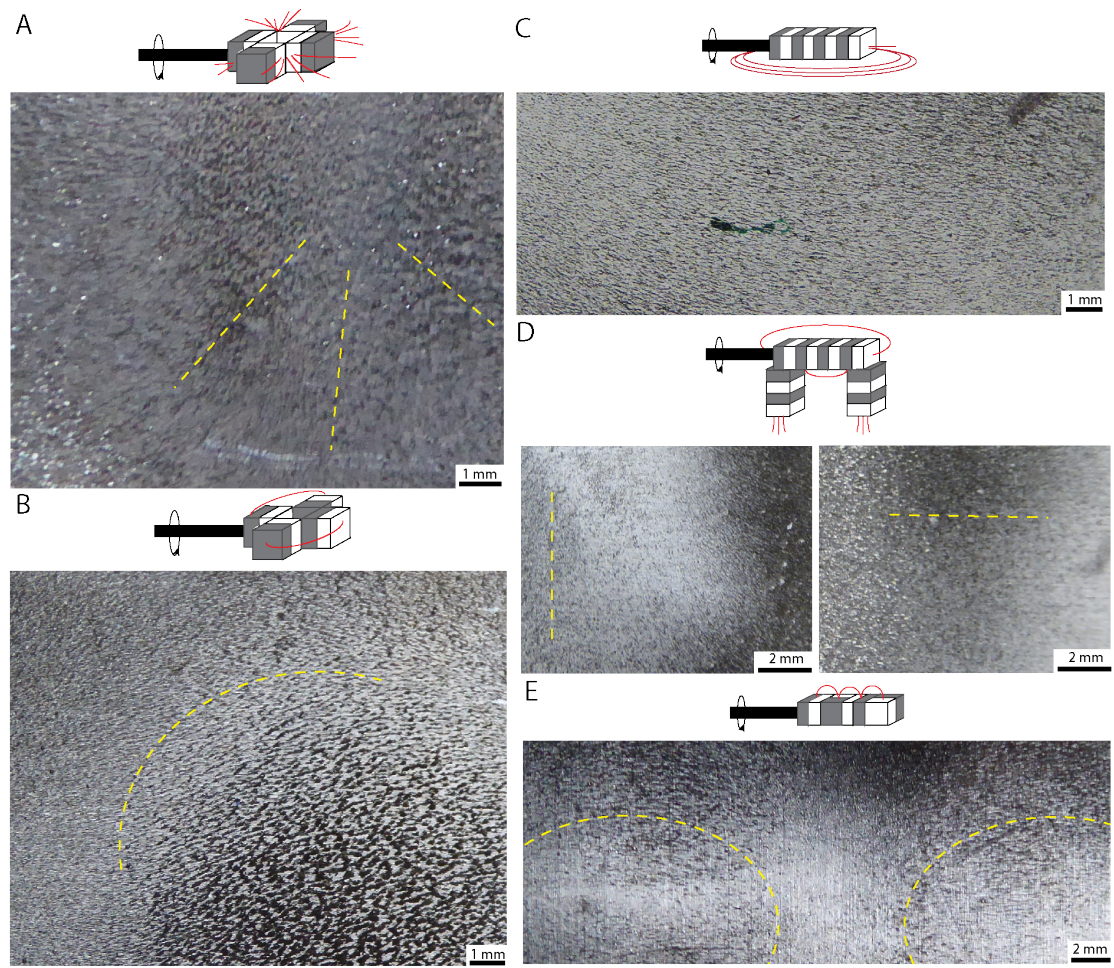

**Figure 7: Alignment patterns in Ni-PDMS obtained using rotating magnet arrays: (A)** radial, **(B)** circular, **(C)** aligned, **(D)** combination of two perpendicular directions, and **(E)** locally curved orientations.

## 5| Conclusions and outlook

In this paper, we presented a practical approach to modulate the in-plane orientation of vertically oriented ferromagnetic Nickel flakes in polymer composites. The approach consists on distributing the flakes in uncured viscous polymeric matrices and to rotate arrays of magnets above the mixture while it is curing. The time for magnetic orientation depends on the flakes magnetic properties, the magnetic field strength, and the viscosity of the matrix and is typically varies from a few seconds to a few minutes.[28,47] The curing method can thus be chosen according to allow for the magnetic orientation. After curing, the flakes retain their orientation. We also show that the in-plane orientation can be predicted from the rotating magnetic field streamlines. This feature is key to further develop a reverse-engineering approach to design complex in-plane orientations in composites according to geometrical, structural, or functional demands. To achieve this, computational modelling using finite element could be used.

Following the flake orientation, the local variation of the particle concentration could also be explored using the same approach. Gradients in magnetic fields indeed lead to magnetophoresis and particle concentration in specific areas. Magnetophoresis and local concentration can be emphasized using particles with strong magnetic susceptibility, or on the contrary, can be reduced with particles with weak magnetic responsiveness. For example, paramagnetic particles or diamagnetic particles with a paramagnetic coating would be less prone to move under magnetic gradients of less than 100 mT, such as those used here. Furthermore, homogeneous distributions of flakes without the chain formation could be achieve using paramagnetic particles too. Indeed, they would have weaker dipole-dipole interactions. In the literature , numerous particle-reinforced composites with magnetic orientation employ paramagnetic particles that do not form chains even at high concentrations and small interparticle distance.[29,48,49] Combining magnetic orientation and magnetic concentration in 3D composites would allow a large degree of design freedom. This is particularly interesting for the fabrication of composites with complex geometries or submitted to complex loadings.

An illustrative application of this design capability could be for realizing bioinspired reinforced composites. Similarly to bone structures where the minerals are oriented and concentrated along the load-bearing directions,[50–52] composites could have their internal microstructure set to withstand specific loadings. This would increase the performance and durability of lightweight materials. Another promising application for locally microstructured composites is for shape-adaptable and magnetically actuated materials for robotics or medicine.[53–55] Bioinspired approaches enabling the distributed orientation of particles to control the local anisotropy have been demonstrated using fibres and 3D printing.[56,57] Utilizing 2D platelets for such materials could increase their morphing capabilities. However, 2D particle orientations cannot be controlled in 3D printing unless combined with magnetic fields. Currently, magnetic 3D printing is time-consuming because each little voxel has to be oriented independently.[36,58] This method thus becomes challenging when applied to larger objects.[59] The use of rotating magnets would solve this problem by being a facile, efficient, and scalable method. Finally, recent development with magnet printing could generate more complex and tuneable magnetic fields.[60–62]

**Associated content**

*Movie S1*

Ni-PDMS composite under a rotating magnetic precessing at an angle of $\theta_{B_0} = 45°$ and a frequency of 0.8 Hz, below the critical frequency for biaxial alignment.

*Movie S2*

Ni-PDMS composite under a rotating magnetic precessing at an angle of $\theta_{B_0} = 45°$ and a frequency of 3.2 Hz, above the critical frequency for biaxial alignment.

**Notes**

The authors declare no competing financial interest.

**Acknowledgements**

The authors acknowledge financial support from the Ministry of Education, Singapore under Grant No. 2019-T1-001-002.